\pdfoutput=1
\documentclass[a4paper,11pt]{article}
\usepackage{amssymb,amsmath,amsthm}
\usepackage[mathscr]{eucal}
\usepackage[pdftex]{hyperref}
\usepackage{url}
\usepackage{graphicx, subfigure}
\usepackage{color}
\usepackage[authoryear,round]{natbib}

\newcommand{\vect}{\mathbf}
\newcommand{\matr}{\mathbf}
\newcommand{\cov} {\operatorname{Cov}}
\newcommand{\cor} {\operatorname{Cor}}

\newcommand{\argmin}{\operatorname{argmin}\;}
\newcommand{\tr}{^T}

\newcommand{\upth}{\mbox{$^{\mathrm{th}}$}}

\newcommand{\txtn}{\textnormal}

\newcommand{\qa}[1]{Eqn.~\txtn(\ref{#1}\txtn)}

\newcommand{\qaa}[1]{\txtn(\ref{#1}\txtn)}

\newcommand{\beq}{\begin{equation}}
\newcommand{\beqa}{\begin{eqnarray}}
\newcommand{\eeq}{\end{equation}}
\newcommand{\eeqa}{\end{eqnarray}}
\newcommand{\bi}{\begin{itemize}}
\newcommand{\ben}{\begin{enumerate}}
\newcommand{\bdes}{\begin{description}}
\newcommand{\edes}{\end{description}}
\newcommand{\ei}{\end{itemize}}
\newcommand{\een}{\end{enumerate}}
\newcommand{\bt}{\begin{tabbing}}
\newcommand{\et}{\end{tabbing}}
\newcommand{\btbl}{\begin{table}}
\newcommand{\etbl}{\end{table}}
\newcommand{\ba}{\begin{array}}
\newcommand{\ea}{\end{array}}

\newcommand{\lp}{\left(}
\newcommand{\rp}{\right)}

\newcommand{\bmt}{\mbox{$\mathbf t$}}
\newcommand{\bmu}{\mbox{$\mathbf u$}}

\begin{document}


\title{Tellipsoid: Exploiting inter-gene correlation for improved detection of differential gene expression}
\author{Keyur Desai\,$^{1}$\footnote{to whom correspondence should be addressed; contact: desaikey@egr.msu.edu}, J.R.~Deller, Jr.\,$^{1}$ and J.~Justin McCormick\,$^2$ \\ $^{1}$Dept. of Electrical and Computer Engineering\\
$^{2}$Dept. of Biochemistry and Molecular Biology \\ Michigan State University, E.~Lansing, MI 48824, USA}

\date{February  11, 2008}

\maketitle
\begin{abstract}

\section*{Motivation:}
Algorithms for differential analysis of microarray data are  vital to modern biomedical research. Their accuracy strongly depends  on   effective  treatment of inter-gene correlation. Correlation is ordinarily accounted for in terms of  its effect on significance cut-offs. In this paper it is shown that correlation can, in fact, be exploited  {to share information across tests}, which, in turn, can increase statistical power.

\section*{Results:}
Vastly and demonstrably improved differential analysis approaches are the result of combining identifiability (the fact that in most microarray data sets, a large proportion of genes can be identified \textit{a priori} as non-differential) with  {optimization} criteria that  incorporate correlation.  As a special case, we develop a method which builds upon the widely used two-sample $t$-statistic based approach and uses the Mahalanobis distance as an optimality criterion. Results on the prostate cancer data of \cite{singh2002gec} suggest that the proposed method outperforms all published  approaches in terms of statistical power.

\section*{Availability:}
The proposed algorithm is  implemented   in MATLAB and in R. The software, called \textit{Tellipsoid}, and relevant data sets are available at \url{www.egr.msu.edu/~desaikey}.
\end{abstract}

\pagestyle{myheadings}
\markboth{Tellipsoid: Exploiting inter-gene correlation}{Submitted to Bioinformatics \qquad \quad Desai \textit{et~al.} }

\section{Introduction}\label{sec.intro}

Detecting differentially expressed genes in the presence of substantial inter-gene correlation is a challenging problem. Research has focused largely on understanding the harmful effects of correlation on the threshold settings demarcating null and non-null genes.  In fact, however, the nominally confounding correlation can be exploited to increase statistical power of microarray studies. This observation has engendered a fruitful  research direction introduced in this paper.

The literature is not devoid of attempts to   develop   more powerful summary statistics  {which exploit correlation},  but such efforts have not produced compelling results.   We posit that the limitations of such developments are due,    at least in part,  to neglecting \emph{identifiability} -- the fact that in most microarray data sets, a large proportion of genes requires no technical effort to be \emph{identified} as explicitly non-differential.

In this paper, we present a new differential analysis method which exploits identifiability through an optimization criterion which incorporates inter-gene correlation. The method builds upon the widely-used two-sample $t$-statistic  approach with  a minimization of the Mahalanobis distance as the optimality criterion. Although this paper  focuses primarily on   two-sample microarray studies,  the framework is  readily generalizable for use in   studies with multiple or continuous covariates, as well as to other multiple comparison applications.

Let us assume that  $m$ genes are measured on $n$ microarrays, under two different experimental conditions, such as control and treatment. Based on the gene expression matrix $X$, we are interested in identifying a small number of genes that are responsible for class distinction. One widely used strategy [e.g.,  \cite{tusher2001sam} and \cite{efron2004lss}] is to represent each gene by its null hypothesis and two-sample $t$-statistic,\footnote{null = non-differential; non-null = differential} say
\begin{eqnarray*}
\mbox{Null hypotheses:}\quad  &H_1,H_2,\ldots,H_m \\
\mbox{Test  statistics:} \quad &t_1,t_2,\ldots,t_m.
\end{eqnarray*}
The magnitudes of the $t$-statistics  establish  a gene-ranking and the $R$ \\ (presumably$\,\ll m$)   genes with the largest $t$-scores are reported as statistically significant discoveries. The investigator can either explicitly supply a $R$ or rely on the \textit{false discovery rate} (FDR) calculations to find a maximal $R$ with the allowable FDR.

The issue of correctly estimating the FDR in the presence of correlation has received much recent attention because highly correlated tests increase the variance of the FDR leading to unreliable results~\citep{owen2005vnf}. As discussed in~\cite{efron2007cal} and \cite{desai2007dnf}, for  ``over powered" $X$'s, there may be  {significantly} fewer tail-area null counts than expected, while for  ``under powered" $X$'s, the situation can worsen with many more tail-area null counts than expected. Importantly though, techniques for correctly estimating the FDR do not change the gene-ranking, but only the size of the reported list.

The  present research was motivated by the notion that, for  ``under powered" $X$'s, it might be possible to  exploit correlation across  $t_i$'s to establish a gene ranking that has better statistical power than the raw $t_i$ based ranking. The  method that resulted from an exploration of this question indeed   seems to improve the statistical power of \emph{all} $X$'s. The proposed method uses, (i) a vector of $t$-statistics ($\vect{t}=[t_1,t_2,\ldots,t_m]\tr$) and (ii) an estimate of the covariance matrix of the vector $\vect t$,  to output   a substantially revised version of $\vect{t}$, denoted   ${\vect{u}}|\vect{t}$, whose corresponding entries can be used to establish an  improved gene-ranking.

The   method is summarized as follows. Let\footnote{No distributional information nor higher order statistics of $\vect t$ are assumed.} $\vect{t} \sim (\vect{u}, \Sigma)$. Now based on the measured value of $\vect{t}$, we can estimate $\vect{u}|\vect{t}$, but while doing so we invoke the  \textit{ zero assumption} (ZA)~\citep{efron2006mem} that the \emph{smallest} $P$(\%) of $t_i$'s are associated with null genes. Based on the ZA, we can set corresponding entries of $\vect{u}$ to zero. For the remaining entries of $\vect{u}$ we obtain \emph{minimum Mahalanobis distance} \citep{mahalanobis1936gds} estimates.

Inter-gene correlation causes $\vect{t}$ to distribute around its center of mass in an hyperellipsoidal manner, and the  Mahalanobis distance is a natural way to measure vector distances in such a distribution.  In fact, the name Tellipsoid is adopted to emphasize the geometric intuition of tracking the center of an ellipsoid. We have done extensive experimentation with both real  and simulated data and found that for a truly null $t_i$ which happens to be in tail-area, the corresponding ${u}_i | \vect{t} $ consistently  tends to   zero (its theoretical value).

Two prior research efforts were particularly useful in  formulating the present approach.  \Citet{storey2007odp}, present  a more general approach to  the notion of sharing information across $t_i$'s, which they describe as ``borrowing strength across the tests,"  for a potential increase in statistical power. \Citet{tibshirani2006csd}, discuss  a new statistic  called the  ``correlation-shared" $t$-statistic, and derive  theoretical bounds on its performance; however, their approach requires strong assumptions regarding the nature of correlation between null and non-null genes which may not hold in many real data sets. The proposed method requires no such assumptions and yet provides considerable increase in power as demonstrated by results reported in Section~\ref{sec.resdis}.

The outline of this paper is as follows. Section~\ref{sec.appro} defines, and obtains  closed form expressions for, the minimum Mahalanobis distance estimates in the $t$-statistic space. Section~\ref{sec.algo} builds on the theory of Section~\ref{sec.appro} to develop the Tellipsoid gene-ranking method. In Section~\ref{sec.resdis} we apply Tellipsoid to real cancer data of \citet{singh2002gec} and compare its accuracy with the state-of-the-art differential analysis tools EDGE~\citep{storey2007odp} and SAM~\citep{tusher2001sam}. Section~\ref{sec.resdis} also discusses the implications of these results. We conclude with Section~\ref{sec.conclusion} summarizing the main ideas.

\section{Approach}\label{sec.appro}

\subsection{Per gene summary statistic}\label{subsec.gen}

Let $X$ be the $m \times n$ matrix of gene expressions, for $m$ genes and $n$ samples. We assume that the samples fall into two groups $k=1$ and $k=2$ and there are $n_k$ samples in group $k$ with $n_1 + n_2 = n$. We start with the standard (unpaired) $t$-statistic:
\begin{equation}
t_i = \frac{\bar{x}_{i;2} - \bar{x}_{i;1}}{s_i},
\label{eq.tstat}
\end{equation}
where $\bar{x}_{i;k}$ is the mean of gene $i$ in group $k$ and $s_i$ is the  pooled within-group standard deviation of gene $i$. If the $i\upth$ gene is indeed null, then we expect $t_i \sim (0,\nu/(\nu-2))$,
where  $\nu$ is the number of degrees of freedom, obtained from either the unpaired $t$-test theory or the permutation null calculations as in \cite{efron2007cal}. If gene $i$ is non-null, then we expect $t_i \sim (u_i,\sigma_i^2)$.   For non-null genes, the values of $u_i$ and $\sigma_i$ depend on the amount of up / down regulation, the number of samples in each group, and $\nu$.

\subsection{The zero assumption}\label{subsec.zero}
Without loss of generality, we may assume that the genes are indexed  so that
\begin{equation}
|t_1| \leq |t_2|\leq \cdots \leq |t_m|.
\label{eq.torder}
\end{equation}
Then, a reasonable way to impose identifiability on null genes is  through the  ZA, namely, that $P$(\%) of the genes -- those with the smallest $t$ statistics --  are null.
\Citet{efron2006mem} discusses the use of the ZA in a variety of differential analysis approaches. It plays a central role in the literature on estimating the proportion of null genes,
as in \citet{pawitan2005bef} and \citet{langaas2005ept}. The ZA is equally crucial for the two-group model approach developed in the Bayesian microarray literature, as in \citet{lee2000irm}, \citet{newton2001dve}, and \citet{efron2001eba}. Furthermore, the claim that the method in \citet{storey2002daf} improves upon the well-known \citet{benjamini1995cfd} FDR procedure (in terms of statistical power) also crucially relies on an adaptive version of the ZA. The use of the ZA is justified in the present situation as long as $P$ is sufficiently small so that the bottom $P$(\%) genes would almost certainly be declared null for reasonable FDR's.

Formally, the ZA is stated as follows:
Let $c$ be the largest integer (gene index) such that $c/m \leq P/100$, denoted
\beq
c=  \lceil{ 0.01 {mP}}\rceil  \label{eq_c},
\eeq
then genes with indices $1,2,\ldots,c$ are assumed null.  Let us partition the set  of  $t$ statistics into those corresponding to genes declared null under the ZA, $\{t_1,t_2,\ldots,t_c\}$, and those for the remaining $m-c$ genes which continue to  \emph{compete} for the non-null designation, $\{t_{c+1},t_{c+2},\ldots,t_m\}$.   For convenience, we introduce the following vector notation,
\begin{align*}
\vect{t} = \left[\ \vect{t}_{1:c}\tr\ \  \vect{t}_{c+1:m}\tr\  \right] \tr= \left[\ \vect{t}_{(0)}\tr\ \  \vect{t}_{(1)}\tr \ \right] \tr.
\end{align*}
Then the random vector $\vect{t}$ is distributed in the following way:
\begin{equation}
\vect{t}\sim (\vect{u},\matr{\Sigma}),
\label{eq.tdist}
\end{equation}
where $\vect{u}$ is the underlying mean vector and $\matr{\Sigma}$ the covariance matrix. The corresponding partitions of $(\vect{u},\matr{\Sigma})$ are denoted
\begin{equation*}
\vect{u} =  \begin{pmatrix}
\vect{u}_{(0)} \\
\vect{u}_{(1)} \\
\end{pmatrix} \mbox{\ \ and\ \ }   \matr{\Sigma} =  \begin{pmatrix}
\matr{\Sigma}_{(00)} & \matr{\Sigma}_{(01)}\\
\matr{\Sigma}_{(10)}   & \matr{\Sigma}_{(11)} \\
\end{pmatrix}.
\end{equation*}
The central hypothesis of this paper is that there is a vector, say  $\vect{u}| \vect{t}$, whose elements represent a reordering of the elements of the $\bmt$, such that gene-ranking represented by $\vect{u}| \vect{t}$ has better statistical power for detecting non-null genes than that based on $\vect{t}$ itself.

(In the present paper, we focus on mainly the second-moment distributional characteristics of $\vect{t}$. However, in fact, if the gene expressions are normally distributed, then, perhaps, $\vect{t}$ is described more accurately by the multivariate Student distribution. Exploitation of this additional structure will be considered in future work.)

\subsection{Estimating $\vect{u}| \vect{t}$}\label{subsec.ugivent}

\newcommand{\ut}{\vect{u}| \vect{t}}

We are interested in obtaining an estimate of the vector mean $\vect{u}$ based on the observation $\vect{t}$. This requires an appropriate metric in the space of  $\bmt$ vectors,
with which to quantify the distance of the observed $\vect{t}$ from the center of mass $\bmu$, say $\mathbf{dist}(\vect{t},\vect{u})$. The $\ell_2$ norm induces a useful metric between  $\vect{t}$ and $\vect{u}$ provided that we first  decorrelate the vector elements as $\matr{\Sigma}^{-1/2}\lp \bmt-\bmu\rp$, thus yielding
 \beqa
\mathbf{dist}(\vect{t},\vect{u}) &=& \sqrt{||\,\matr{\Sigma}^{-1/2}\lp \bmt-\bmu\rp\,||^2}\nonumber\\
&=&
\sqrt{(\vect{t} - \vect{u})^T\matr{\Sigma}^{-1}(\vect{t} - \vect{u})}.
\label{eq.mahala}
\eeqa
This weighted Euclidean distance is sometimes called the \textit{Mahalanobis distance} in the pattern classification literature as in \cite{dellerjr1993dtp}, \cite{devijver1982prs}, etc.

 We can relate $\vect{t}$ to $\vect{u}$ through the Mahalanobis distance but while doing so we invoke the ZA, which, in turn, implies that the first $c$ entries of $\vect{u}$ are zero. This yields the   estimate
\beqa  
\label{eq.ustar}
\lefteqn{\vect{u}^*  =  \begin{pmatrix}
\vect{0} \\
\vect{u}_{(1)}^* \\
\end{pmatrix},\quad \textrm{where}\quad
\vect{u}_{(1)}^* =  \underset{\vect{u}_{(1)} \in \mathbb{R}^{m-c}}{\argmin} } \\  &&\begin{pmatrix}  \vect{t}_{(0)} - \vect{0}  \\   \vect{t}_{(1)} - \vect{u}_{(1)} \\ \end{pmatrix}\tr
\begin{pmatrix}
\matr{\Sigma}_{(00)} & \matr{\Sigma}_{(01)}\\
\matr{\Sigma}_{(10)}   & \matr{\Sigma}_{(11)} \\
\end{pmatrix} ^{-1}
\begin{pmatrix}  \vect{t}_{(0)} - \vect{0}  \\   \vect{t}_{(1)} - \vect{u}_{(1)} \\ \end{pmatrix}. \nonumber
\eeqa 
In effect, $\vect{u}_{(1)}^*$ combines the identifiability information based on the ZA with the information about the covariance structure of $\vect{t}$ which too can be obtained from the measured $X$ itself.
Notably the optimization in \qa{eq.ustar} enjoys closed form solution:
\begin{equation}
\vect{u}_{(1)}^* =  \vect{t}_{(1)} - \matr{\Sigma}_{(10)}\matr{\Sigma}_{(00)}^{-1}\vect{t}_{(0)}.
\label{eq.sol}
\end{equation}
The derivation leading from \qa{eq.ustar} to \qa{eq.sol} is provided in the Appendix.

\newcommand{\ip}{{i^\prime}}
\subsection{Estimating $\textrm{Cov}(t_i,t_\ip)$ }\label{subsec.cov}

To  estimate the required entries of $\vect{\Sigma}$, we make several  observations. The validity of these observations can be established through computer simulations using  the MATLAB script \textsf{testtcov.m}  available with the Tellipsoid software. (Equation~\qaa{eq.sol} does not require the covariance between two non-null $t_i$'s.)
\smallskip

\noindent\textbf{Observation 1.}
If genes $i$ and $\ip$ both are null, then
\begin{equation}
\cov(t_i,t_\ip) \approx \cor(x_i,x_\ip)\,\frac{\nu}{\nu-2}
\label{eq.tcovnn}
\end{equation}
\Citet{efron2007cal} and \Citet{owen2005vnf} use this observation for their conditional FDR calculations. This observation maybe intuitive to the reader from \qa{eq.tstat} itself, or it is easily   verified through a computer simulation.
\smallskip

\noindent\textbf{Observation 2.}
Similarly, if the gene $i$ is null and $\ip$ non-null  (or conversely), then
\begin{equation}
 \cov(t_i,t_\ip) \approx \frac{n_2\cor(x_{i;1},x_{\ip;1}) + n_1\cor( x_{i;2},x_{\ip;2})}{n_1 + n_2}\frac{\nu}{\nu-2},
\label{eq.tcovn-nn}
\end{equation}
where $\cor(x_{i;k},x_{\ip;k})$ denotes the correlation between gene $i$ and $\ip$ within group $k$.  Equation~\qaa{eq.tcovn-nn} accommodates the possibility that the correlation between a null and a non-null gene may change across groups.
If this does not occur, then \qa{eq.tcovn-nn} reduces to \qa{eq.tcovnn}.
\smallskip

\noindent\textbf{Observation 3.}
Furthermore, if $n_1 \approx n_2$ (which is  the case for most microarray studies), then \qa{eq.tcovn-nn} simplifies to:
\begin{equation}
 \cov(t_i,t_\ip) \approx \frac{\cor(x_{i;1},x_{\ip;1}) + \cor( x_{i;2},x_{\ip;2})}{2}\frac{\nu}{\nu-2}.
 \label{eq.tcovnnn}
\end{equation}

Equations~\qaa{eq.tcovnn} and \qaa{eq.tcovnnn} suggest that we may use sample correlation to estimate $\cov(t_i,t_\ip)$:
\begin{equation}
\widehat{\cov}(t_i,t_\ip) \propto \frac{\sum_j \tilde{x}_{ij}\tilde{x}_{\ip j}}{\sqrt{\left(\sum_j \tilde{x}_{ij}^2\right) \left(\sum_j \tilde{x}_{\ip j}^2\right)}},
\label{eq.hatcov}
\end{equation}
where $\tilde{x}_{ij}$ denotes the expression level of the $i\upth$ gene measured on the $j\upth$ microarray after subtracting  the average response within each treatment group. The scale factors cancel in the  terms $\matr{\Sigma}_{(10)}$ and $\matr{\Sigma}_{(00)}^{-1}$, so that estimating  $\nu$/($\nu$-2) [see \qa{eq.tcovnn}] is not required.

\subsection{Tellipsoid equation}\label{subsec.implementation}

In light of \qa{eq.hatcov}, \qa{eq.sol} takes the practical   form
\begin{equation}
\widehat{\vect{u}}_{(1)}^* =  \vect{t}_{(1)} - \widehat{\matr{C}}_{(10)}\widehat{\matr{C}}_{(00)}^{-1}\vect{t}_{(0)} ,
\label{eq.solprac}
\end{equation}
where $\widehat{\matr{C}}$ is the sample correlation matrix of the gene expression matrix $\tilde{X}$ (after removing the treatment effects). In most cases computing the full matrix inverse ($\widehat{\matr{C}}_{(00)}^{-1}$) is not necessary and solving the term $\widehat{\matr{C}}_{(00)}^{-1}\vect{t}_{(0)}$ through some efficient linear solver  reduces the computation considerably. (See Subsection~\ref{subsec.compute} for detail).

\section{Algorithm}\label{sec.algo}

\subsection{Tellipsoid}\label{subsec.tellipsoid}

This section outlines a self-contained differential analysis algorithm based on the ideas discussed in Section~\ref{sec.appro}. Its name Tellipsoid was coined to emphasize the geometric intuition of tracking the center of an ellipsoid.

Tellipsoid takes gene expression matrix $X$ and   provides a specified number, $R$, of the most differentially expressed genes. In principle, the ranking is based on the set $\{u_i^*\}$  from \qa{eq.ustar}. In practice, we rely on the estimates $\{\hat{u}_i^*\}$  from \qa{eq.solprac}.

The algorithm begins by reindexing the genes based on their two-sample $t$-statistics [\qa{eq.torder}]. Then, based on the ZA, the first $c$   genes are identified as null, as specified in \qa{eq_c}. By default, $P$ is set to 50(\%). Although the choice 50\% is somewhat arbitrary, this fraction has worked well empirically in the data sets tested.  Future research may yield more rigorous methods for choosing $P$.

In order to nullify any genuine treatment differences, $X$ is converted to $\widetilde{X}$ by subtracting  each gene's average response within each treatment group. The sample correlation matrix $\widehat{\matr{C}}$ of $\widetilde{X}$ is computed subsequently. The crucial step is to compute $\widehat{\vect{u}}_{(1)}^*$ based on \qa{eq.solprac}. The elements of $\widehat{\vect{u}}^* = [\vect{0}_{c\times 1}\tr\, (\widehat{\vect{u}}_{(1)}^*)\tr]\tr$ determine the gene-ranking: A gene with bigger $|\widehat{u}_i^*|$ is assigned a higher rank. The first $R$ genes are reported as top $R$ statistical discoveries.

\subsection{Numerical stability}\label{subsec.numeric}
Because the number of samples $n$ is often less than the number of genes $m$, the sample correlation matrix turns out to be singular and hence non-invertible. Therefore, we add a very small correction term (= $10^{-10}$) to its diagonal entries to make it nonsingular, and in effect, invertible. After this correction, Tellipsoid shows excellent numerical stability.

\subsection{Computational complexity}\label{subsec.compute}
Equation~\qaa{eq.solprac} involves matrix inversion, which, if performed in a naive way, could be a prohibitive operation, since microarray data sets may have several tens of thousand genes. Indeed, solving the term $\widehat{\matr{C}}_{(00)}^{-1}\vect{t}_{(0)}$ as a system of simultaneous linear equations  ($\widehat{\matr{C}}_{(00)}\vect{x}=\vect{t}_{(0)}$) is much faster than explicitly computing $\widehat{\matr{C}}_{(00)}^{-1}$. In particular, we can employ the Cholesky decomposition to exploit the fact that the matrix $\widehat{\matr{C}}_{(00)}$ is symmetric and positive definite. MATLAB implementation of Tellipsoid uses the in-built \texttt{linslove} with appropriate settings, which, in turn, uses highly optimized routines of LAPACK (Linear Algebra PACKage---\url{http://www.netlib.org/lapack/}).

For the Prostate data (used in Section~\ref{sec.resdis}) with $12625$ genes and $102$ samples, Tellipsoid, running on a computer with a 2.2~GHz dual-core AMD Opteron processor with 8~GB of RAM and MATLAB version R2006b, requires just under 40 seconds to report the final gene list. For the same settings, the implementation with explicit matrix inversions takes $\sim10$ minutes.

\subsection{Summary of Tellipsoid}

\hrulefill \\
\textbf{Tellipsoid:} An improved differential analysis algorithm \\
\textbf{Input:}  $X=$ Labeled $m\times n$ gene expression matrix; $R=$ Size of gene list\\
\textbf{Output:} The gene list containing top $R$ differentially expressed genes
\begin{enumerate}
  \item Calculate two-sample (unpaired) $t$-statistics: $t_i = (\bar{x}_{i;2} - \bar{x}_{i;1})/{s_i}$
  \item Reindex genes such that $|t_1| \leq |t_2|\leq \cdots \leq |t_m|$
  \item Gather first $c$ = $\lceil{ 0.01 {mP}}\rceil$ $t_i$'s in a vector $\vect{t}_{(0)}$; By default $P$=50(\%)
  \item Convert $X$ to $\widetilde{X}$ by subtracting each gene's average response within each treatment group
  \item Compute $\widehat{\matr{C}}$ = the sample correlation matrix of $\widetilde{X}$
  \item Find $\widehat{\vect{u}}^*$ = $[\vect{0}_{c\times 1}\tr\, \left(\widehat{\vect{u}}_{(1)}^*\right)\tr]\tr$, where $\widehat{\vect{u}}_{(1)}^*$ = $\vect{t}_{(1)} - \widehat{\matr{C}}_{(10)}\widehat{\matr{C}}_{(00)}^{-1}\vect{t}_{(0)}$
  \item Determine gene-ranking: Assign a gene with bigger $|\widehat{u}_i^*|$ a higher rank
  \item Report top $R$ genes as statistical discoveries
\end{enumerate}
\hrulefill

\newcommand{\subfigwidthb}{0.49}
\newcommand{\subfigwidths}{0.49}
\section{Results and Discussion}\label{sec.resdis}
We compare Tellipsoid with two of the leading techniques, SAM [Significance Analysis of Microarrays~\citep{tusher2001sam}] and EDGE [Extraction and analysis of Differential Gene Expression~\citep{leek2006eea}]. SAM adds a small exchangeability factor $s_0$ to the pooled sample variance when computing the two-sample $t$-statistic: $d_i = (\bar{x}_{i;2} - \bar{x}_{i;1})/(s_i + s_0)$; whereas EDGE is based on a general framework for sharing information across tests (see \citet{storey2007odp}). EDGE is reported to show substantial improvement (in terms of statistical power) over five of the leading techniques including SAM \citep{storey2007odp}. The other four are: (i) $t$/$F$--test of \citet{kerr2000avg} and \citet{dudoit2002smi}; (ii) Shrunken $t$/$F$--test of \Citet{cui2005ist}; (iii) The empirical Bayes local FDR of \citet{efron2001eba}; (iv) The \textit{a posteriori} probability approach of \citet{lonnstedt2002rmd}. It is noteworthy that Tellipsoid shows a major improvement over EDGE itself.

To determine the performance quality of various techniques, we focus primarily on the empirical FDR in the reported gene list: Empirical FDR = $\textrm{NoFP}/R$, where NoFP = the number of false positives. Broadly speaking, smaller the FDR better the technique.

\subsection{Prostate data}
\citep{singh2002gec} studied $m=12625$ genes on $n=102$ oligonucleotide microarrays, comparing $n_1=50$ healthy males with $n_2=52$ prostate cancer patients. The purpose of their study was to identify genes that might anticipate the clinical behavior of Prostate cancer. We downloaded the .CEL files from \url{http://www-genome.wi.mit.edu/MPR/prostate}. The software RMAExpress \citep{irizarry2003sag} was used to obtain high quality gene expressions from these .CEL files. We let RMAExpress apply its in-built background adjustment, however, the quantile normalization was skipped. Each gene was represented in the final expression matrix $X$ by the logarithm (base 10) of its expression level. Taking the $\log$ is thought to increase normality and stabilize across group standard deviations \citep{tsai2003tde}.

\subsubsection{Data with known results}\label{subsec.dataanswers}
Algorithm testing required  an expression matrix $X$ with the knowledge of truly non-differential genes. At the same time,  we wanted the  inter-gene correlation in $X$  to resemble that in the real microarray data. These two seemingly conflicting requirements  were satisfied concurrently  by row standardizing a real $X$. The prostate cancer matrix $X$  was transformed to $\underline{\widetilde{X}}$ by subtracting each gene's average response within each treatment group, and by normalizing  within group sample mean squares.  That is,  for each group $k\in\{1,2\}$, $(1/n_k) \sum_j \underline{\widetilde{X}}_{ij} = 0$ and $(1/n_k) \sum_j \underline{\widetilde{X}}_{ij}^2 = 1$. Here, the sum runs over corresponding $n_k$ samples only. With this transformation,  all genes have equal energy and yet the same within group inter-gene correlation structure as the original $X$.\footnote{\textbf{Note.} Normalizing within group sample mean squares to unity is not implemented in the Tellipsoid algorithm.}

To generate a test data set from $\underline{\widetilde{X}}$, its 102 columns were randomly divided  into groups of 50 (=$n_1$) and 52 (=$n_2$). Next $m_u$ ($m_d$)   genes were randomly chosen for up (down) regulation  by adding a positive (negative) offset $x_u$ ($x_d$) to the corresponding entries in group 2. Various choices of $(m_u,m_d,x_u,x_d)$ were tested to represent a range of differential analysis scenarios encountered in practice.

\subsubsection{Results for prostate data}
Two cases were studied.   In the first, the proportion of truly differential genes,  say $p_1$, was taken to be relatively small: $p_1$ $\sim$ 0.01--0.05. The second case employed a larger $p_1$ $\sim$ 0.1. The former simulates   microarray studies seeking genes that distinguish subtypes of cancer, diabetes, etc., whereas the latter resembles studies comparing healthy versus diseased cell activity.

Results were obtained using the subroutines \texttt{samr.r} from the package ``samr" and \texttt{statex.r} from the package ``edge."  Both routines compute their native gene summary statistics given the matrix  $X$ and corresponding column labels.  These statistics, in turn, can be used to determine top $R$ genes.

\begin{figure}[ht]
\centering
\subfigure[]{\label{fig.Ttl300R300}\includegraphics[width=\subfigwidthb\columnwidth]{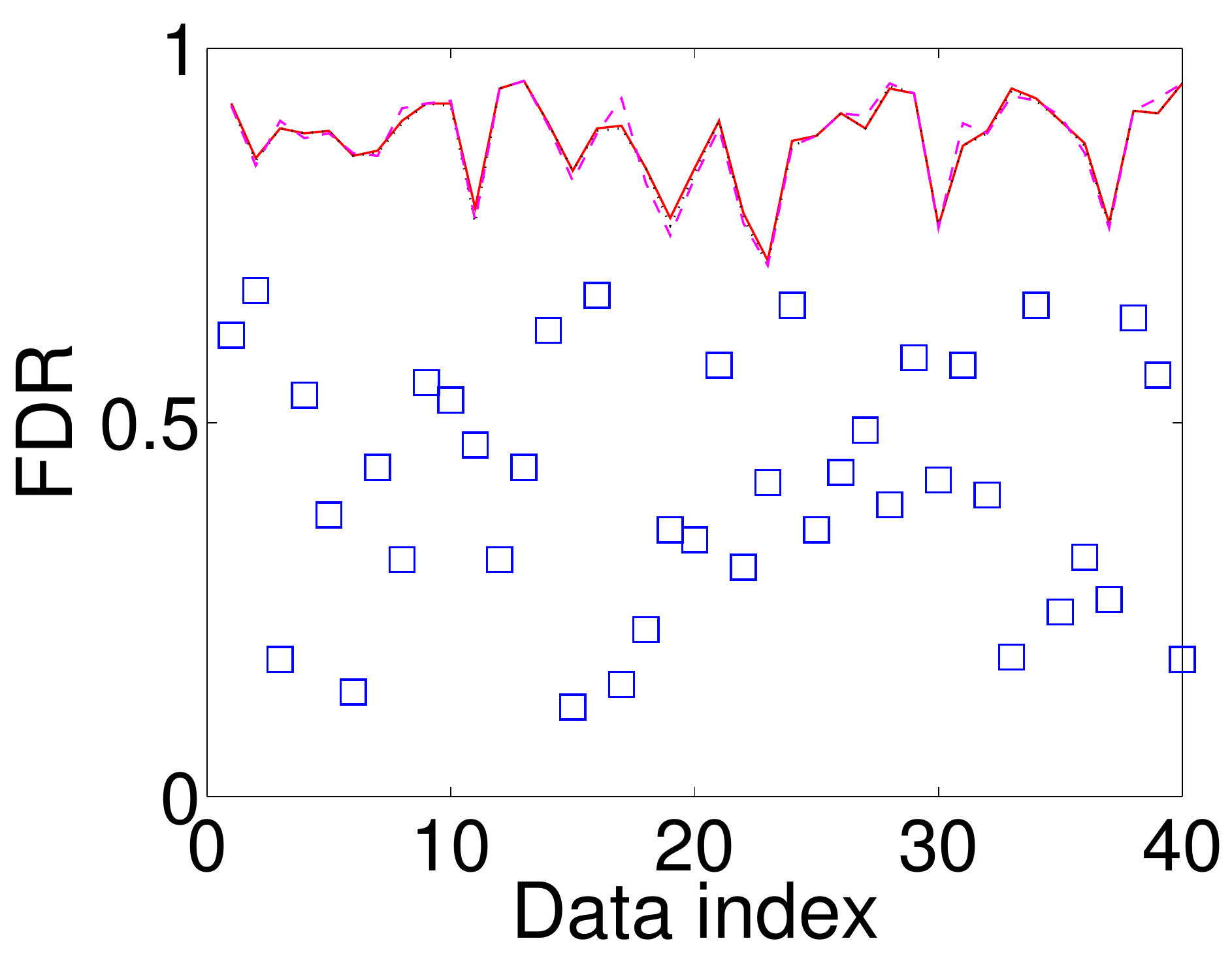}}
\subfigure[]{\label{fig.Ttl300R100}\includegraphics[width=\subfigwidthb\columnwidth]{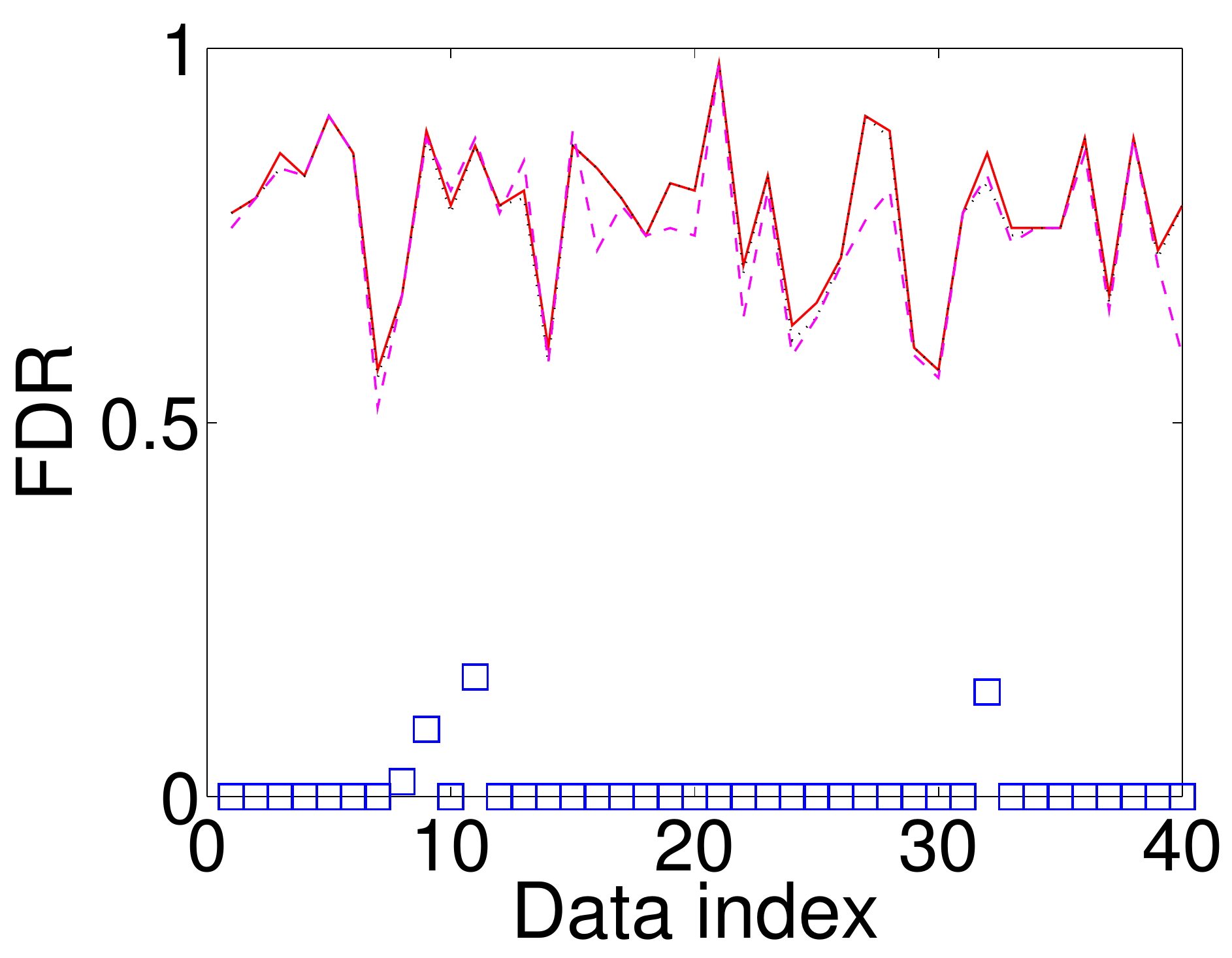}}
\caption{FDRs for Case 1. The number of truly differential gene is 300. Panel (a) $R$=300; Panel (b) $R$=100. Tellipsoid offers better detection. Square $(\square)$ marker = Tellipsoid. Lines: solid = raw $t$-statistic; dotted = SAM; dashed = EDGE.}
\label{fig.Ttl300}
\end{figure}

\smallskip
\noindent\textbf{Case 1} [$p_1 \approx 0.025$, $m_u$=200, $m_d$=100, $x_u$=0.1, and $x_d$=-0.1].
Figure~\ref{fig.Ttl300R300} shows plots of the FDRs for 40 different data sets with the size of the reported list, $R$=$300$. A large value of $R$ coincides with an attempt  to extract as many differential expressions as possible, a desired goal  especially in microarray studies performed to identify genes that are to be explored further  --  experimentally   or computationally --  to gain better understanding of  underlying gene networks. Since the differential signal $x_u$=0.1 and $x_d$=-0.1 is rather weak, recovering a good list is not easy as evident from the results -- among all methods only Tellipsoid achieved  sufficiently low  FDRs to rescue a few $X$'s.

\begin{figure}[h]
\centering
\subfigure[]{\label{fig.Ttl1200R1200}\includegraphics[width=\subfigwidths\columnwidth]{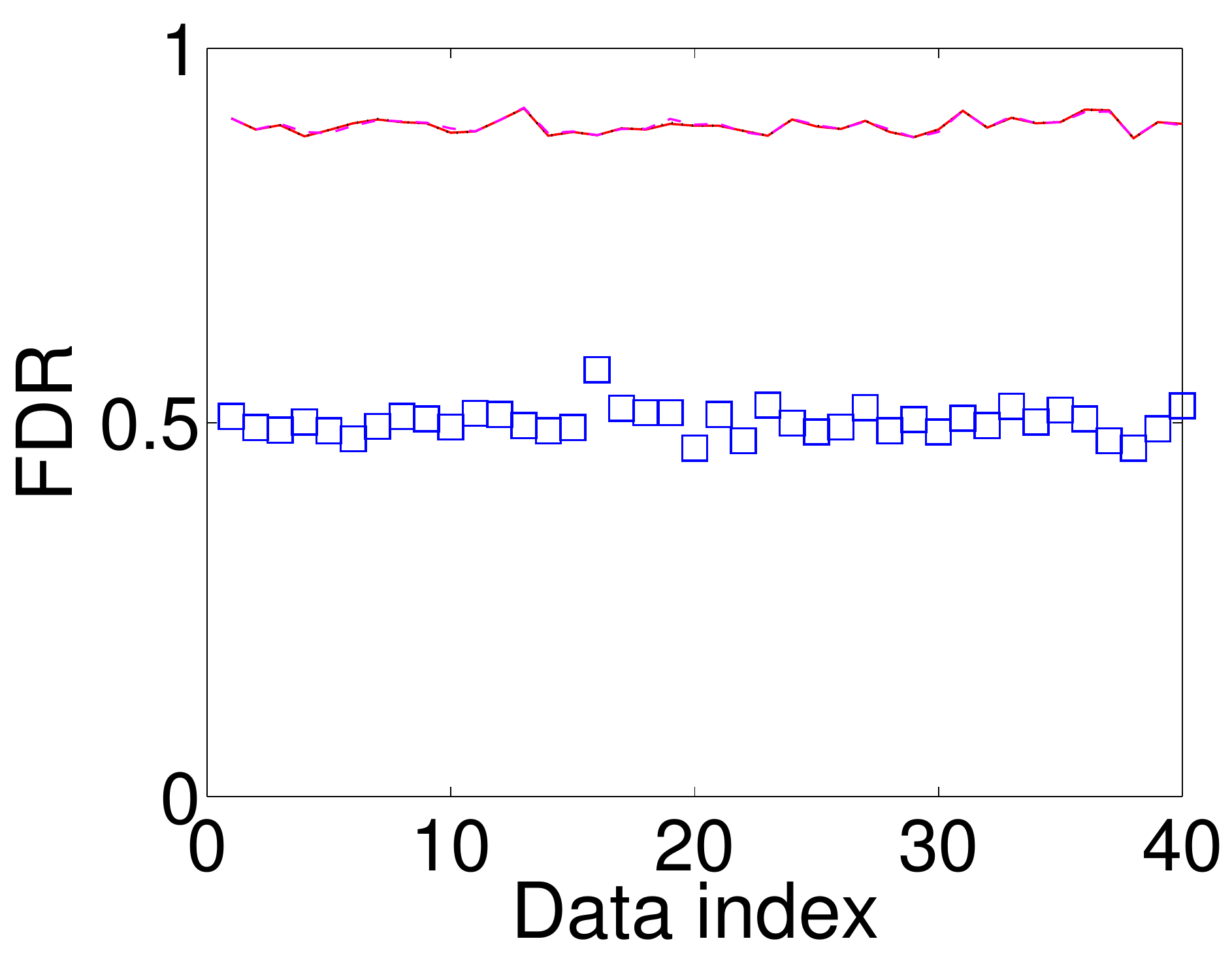}}
\subfigure[]{\label{fig.Ttl1200R1200}\includegraphics[width=\subfigwidths\columnwidth]{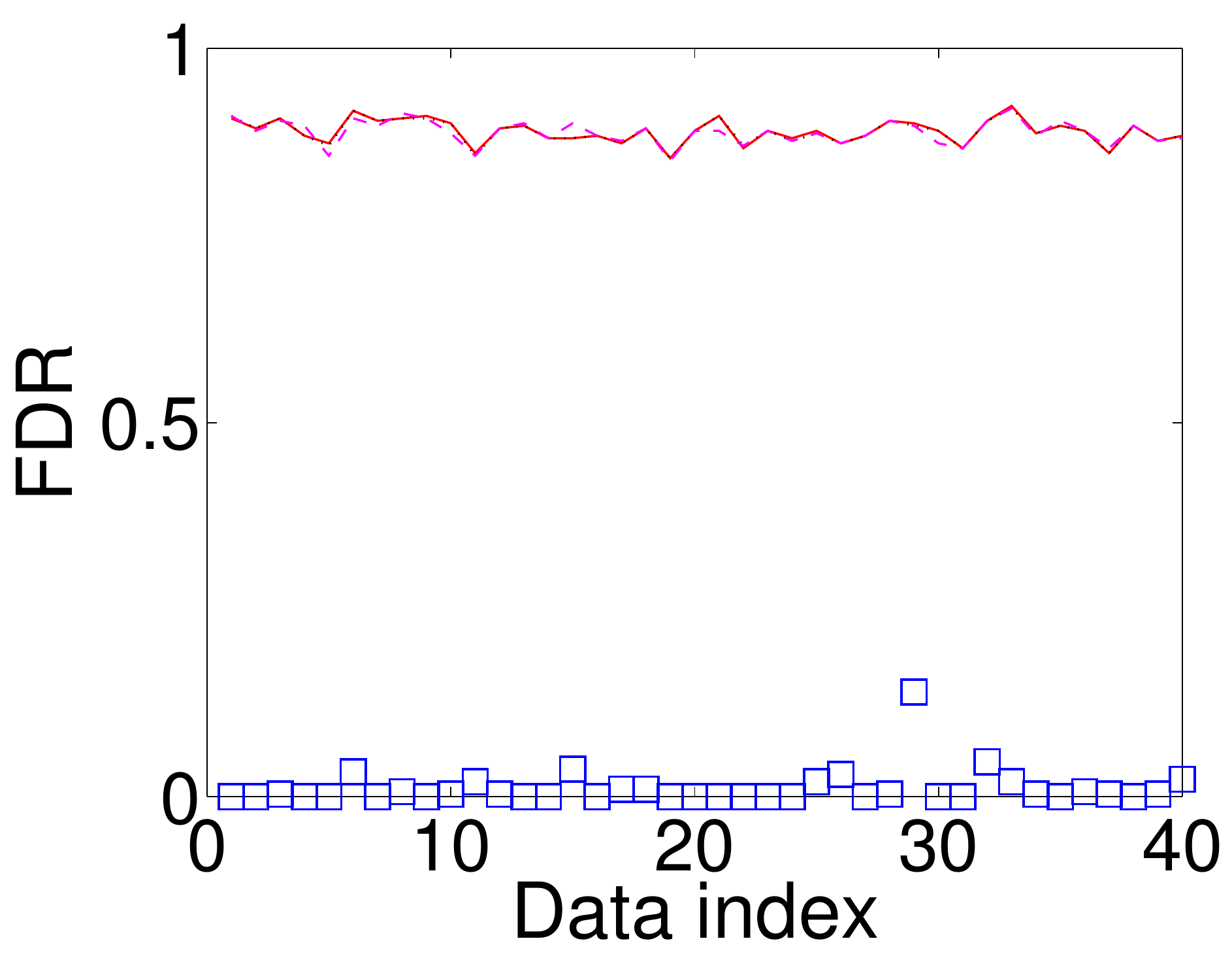}}
\caption{FDRs for Case 2a. The number of truly differential gene is 1200. Panel (a) $R$=1200; Panel (b) $R$=300. Tellipsoid performs much better. Square $(\square)$ marker = Tellipsoid. Lines: solid = raw $t$-statistic; dotted = SAM; dashed = EDGE.}
\label{fig.Ttl1200}
\end{figure}

Figure~\ref{fig.Ttl300R100} presents results for  $R$=100. A smaller $R$ would be chosen to   identify high-quality class distinguishing features for gene-expression-profiling-based clinical diagnosis and prognosis, where the goal is to build accurate classifiers and predictors. Whereas \cite{singh2002gec} build a classifier around only 16 of 12625 features, they do discuss the need to include as many reliable features as possible. Remarkably, for 37 out of 40 $X$ matrices, Tellipsoid reports gene lists with no false discoveries at all, while the other techniques fail to obtain a single gene list with an FDR$<$0.5.
\smallskip

\noindent\textbf{Case 2a} [$p_1 \approx 0.1$, $m_u$=600, $m_d$=600, $x_u$=0.02, and $x_d$=-0.02].
In this set of experiments,  $p_1$ is increased, but  the differential signal is reduced. This situation also proves to be challenging for the existing techniques. However,  Tellipsoid   provides the FDR of $\sim0.5$ for $R$=1200, and, again for $R$=300, while it reports most gene lists with no false discoveries at all.
\smallskip

\begin{figure}[h]
\centering
\subfigure[]{\label{fig.Ttl1200R1200nt20}\includegraphics[width=\subfigwidths\columnwidth]{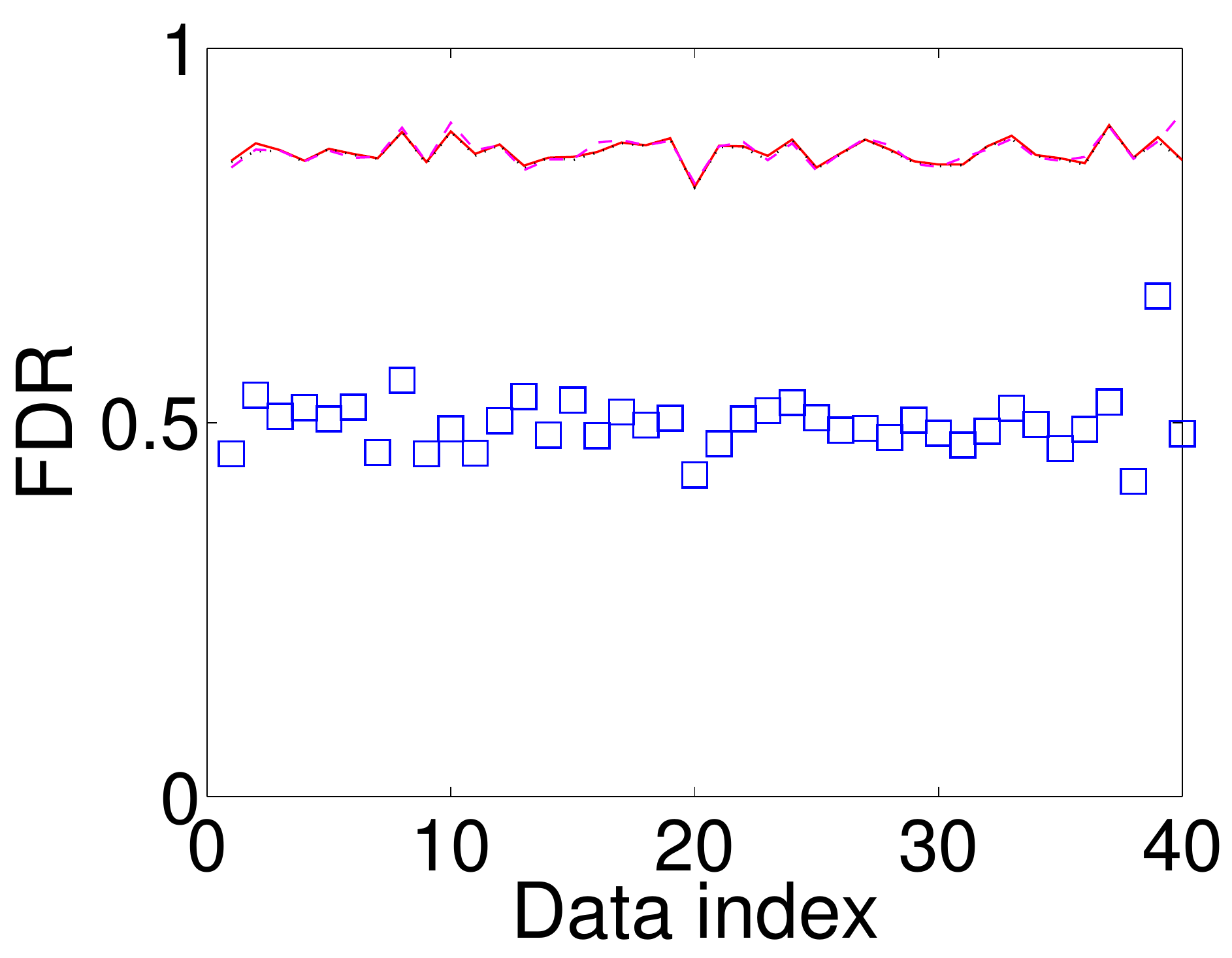}}
\subfigure[]{\label{fig.Ttl1200R1200nt20}\includegraphics[width=\subfigwidths\columnwidth]{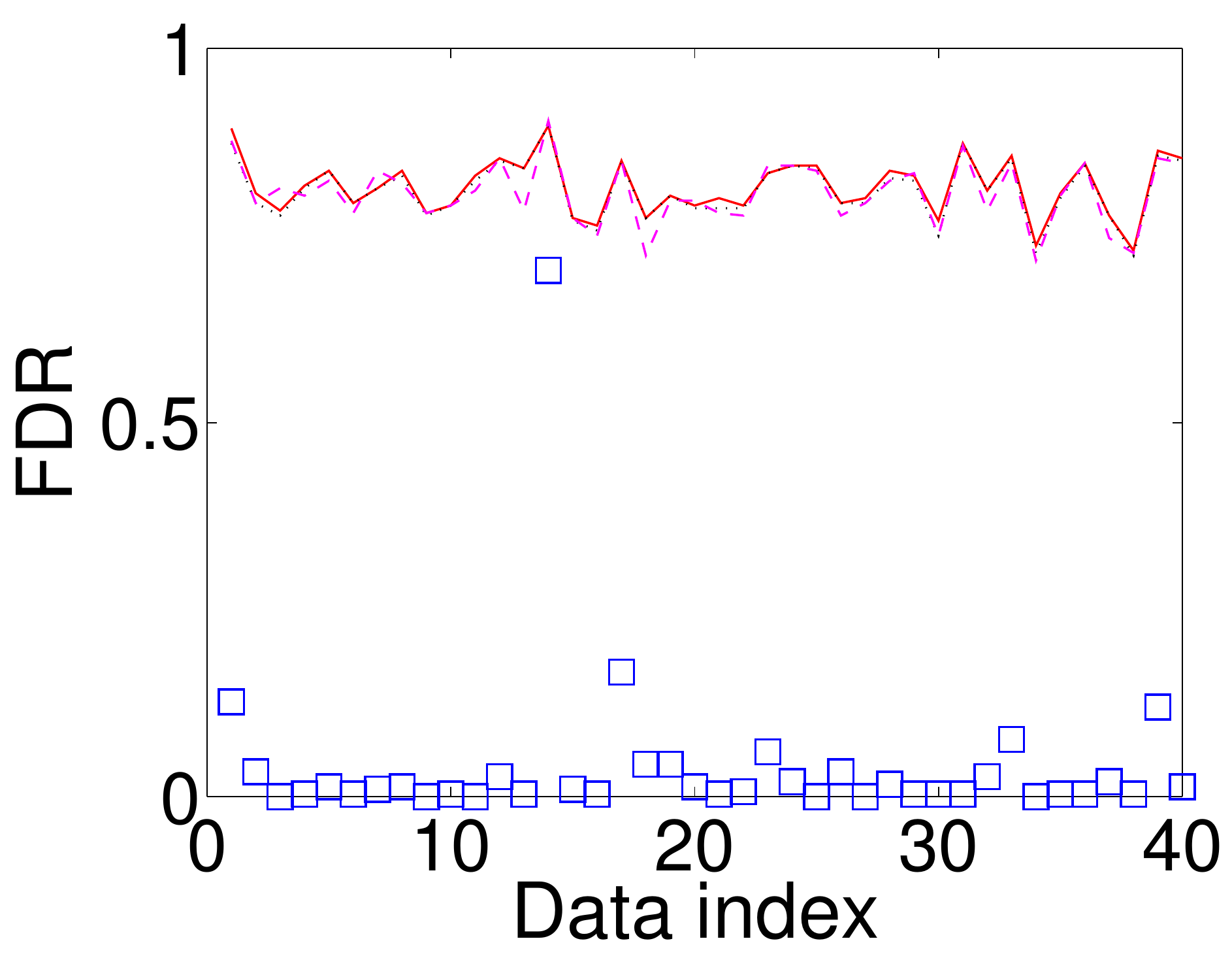}}
\caption{FDRs for Case 2b. Panel (a) $R$=1200; Panel (b) $R$=300. The sample size is smaller than that in Cases 1 $\&$ 2a and yet Tellipsoid performs well. Square $(\square)$ marker = Tellipsoid. Lines: solid = raw $t$-statistic; dotted = SAM; dashed = EDGE.}
\label{fig.Ttl_1200_nt20}
\end{figure}

\noindent\textbf{Case 2b} [$p_1\approx$0.1, $m_u$=600, $m_d$=600, $x_u$=0.1, and $x_d$=-0.1].
This subcase is designed  to assess  the effects of small sample sizes on   performance.  $n_1$ and $n_2$ are both reduced  to 20. We randomly chose 20 columns per group from the original prostate cancer $X$, and then applied the data generation process (including row standardization) detailed in Subsection~\ref{subsec.dataanswers}. Reduction in the number of samples is compensated by increase in the differential signal. The FDRs for Tellipsoid, Fig.~\ref{fig.Ttl_1200_nt20}, are excellent suggesting that Tellipsoid increases power of small sample data sets too.

\subsection{Simulated data}
Before devising the test data setup of Subsection~\ref{subsec.dataanswers}, Tellipsoid was tested on several simulated data sets. Below we discuss some simulation results that  shed further light on the small sample behavior.

\begin{figure}[h]
\centering
\subfigure[]{\label{fig.Ttl100R100nt10}\includegraphics[width=\subfigwidths\columnwidth]{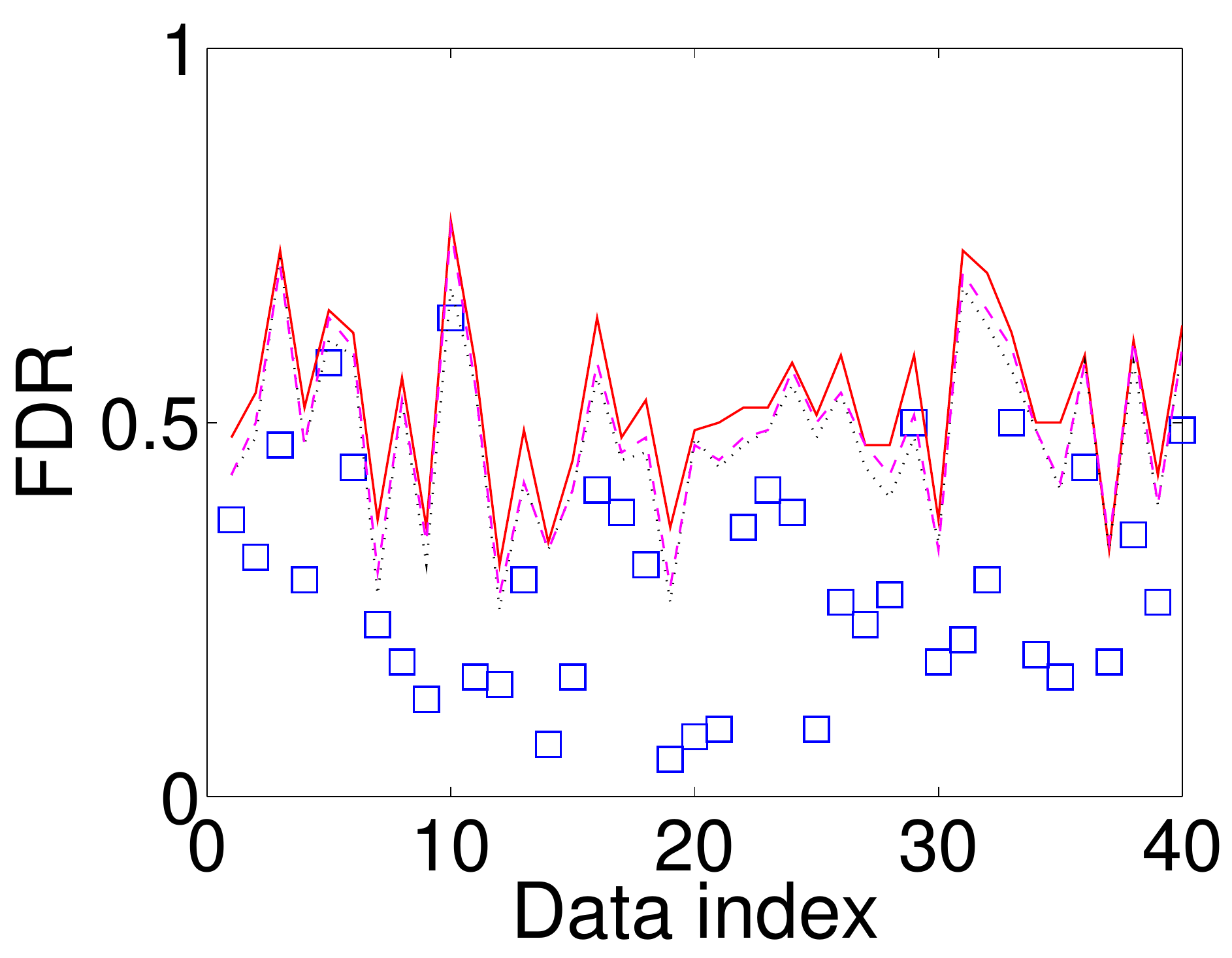}}
\subfigure[]{\label{fig.Ttl100R100nt10}\includegraphics[width=\subfigwidths\columnwidth]{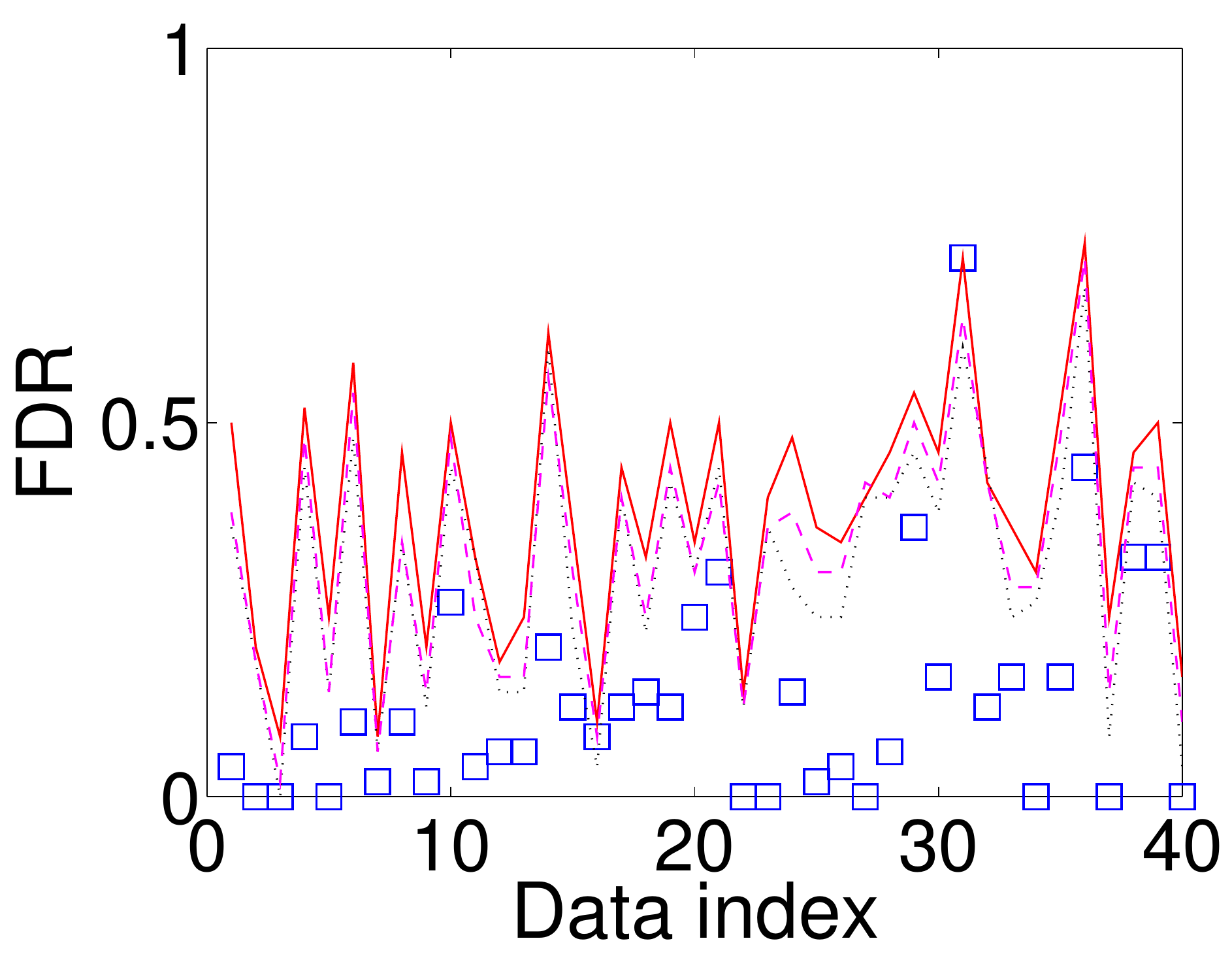}}
\caption{FDRs for simulated data. Panel (a) $R$=100; Panel (b) $R$=50. (Small) sample sizes: $n_1$=10, $n_2$=10. Yet, Tellipsoid performs better than the rest. Square $(\square)$ marker = Tellipsoid. Lines: solid = raw $t$-statistic; dotted = SAM; dashed = EDGE.}
\label{fig.Ttl_100_nt10_sim}
\end{figure}

Let us denote by  $X_{(i)}$ the $i\upth$ column of a  simulated expression matrix $X$.  We assume that the random vector $X_{(i)}$ is multivariate Gaussian with mean $\vect 0$ and covariance matrix $\matr W$.  Each such column represents $m$=3226 genes with a  covariance matrix $\matr{W}$ that introduces roughly the same amount of correlation as found in the BRCA data of \cite{hedenfalk2001gep}. We choose $m_u =50,    m_d =50,    x_u =1,    x_d =-1,   n_1 =10$, and $n_2=10$. Figure~\ref{fig.Ttl_100_nt10_sim} shows plots of  the FDRs for $R$=50 and $R$=100. Table~\ref{tab.tiui} shows results for some $X$ realization from Fig.~\ref{fig.Ttl100R100nt10}. Shown are the top 100 values of $\hat{u}^*_i$  and each corresponding  original $t_i$  with concomitant rank. With smaller $n$, preeminence of Tellipsoid with respect to existing techniques scales down a bit.  Nevertheless, for $R$=50 case, for 25 out of 40 simulated $X$ realizations, Tellipsoid achieves a low   FDR of $\sim$ 0.1 or less.

\begin{table}
  \centering
  \newcommand{\nullc}[1]{\textcolor[rgb]{0.00,0.00,0.00}{{\textsf{\textbf{#1}}}}}
\footnotesize
\begin{tabular}{rrrr|rrrr}
 \hline \hline 
\multicolumn{4}{c}{$1$--$50$}&\multicolumn{4}{c}{$51$--$100$}\\ \hline 
$\hat{u}^*_i$ rank &$\hat{u}^*_i\,$ &$t_i\,$ &$t_i$ rank& $\hat{u}^*_i$ rank &$\hat{u}^*_i\,$ &$t_i\,$ &$t_i$ rank \\ \hline 
1 &4.22 &5.87 &1\, &\nullc{51} &\nullc{2.18} &\nullc{3.45} &\nullc{23}\, \\ 
2 &-4.17 &-5.55 &2\, &\nullc{52} &\nullc{2.17} &\nullc{3.05} &\nullc{42}\, \\ 
3 &-3.93 &-4.26 &5\, &53 &2.16 &2.80 &82\, \\ 
4 &-3.74 &-4.12 &7\, &54 &-2.15 &-2.57 &122\, \\ 
5 &-3.58 &-4.49 &4\, &55 &2.15 &1.96 &357\, \\ 
6 &-3.49 &-3.34 &28\, &56 &-2.14 &-1.47 &751\, \\ 
7 &-3.45 &-4.25 &6\, &57 &2.13 &2.25 &229\, \\ 
8 &3.35 &3.87 &10\, &58 &2.13 &1.77 &486\, \\ 
9 &-3.33 &-3.20 &35\, &59 &2.13 &1.44 &785\, \\ 
10 &3.33 &3.77 &13\, &60 &2.12 &2.14 &273\, \\ 
11 &3.25 &3.42 &25\, &61 &-2.11 &-1.48 &744\, \\ 
12 &-3.16 &-2.18 &260\, &\nullc{62} &\nullc{2.10} &\nullc{1.80} &\nullc{453}\, \\ 
13 &3.14 &4.54 &3\, &63 &2.09 &2.60 &114\, \\ 
14 &3.10 &2.87 &65\, &64 &-2.09 &-2.05 &312\, \\ 
15 &-3.08 &-3.54 &17\, &65 &2.09 &2.70 &96\, \\ 
16 &-3.07 &-2.80 &80\, &66 &2.09 &2.23 &237\, \\ 
17 &3.06 &3.49 &20\, &67 &2.08 &2.34 &188\, \\ 
18 &3.02 &2.29 &213\, &68 &-2.08 &-2.24 &232\, \\ 
19 &-2.99 &-3.34 &27\, &69 &-2.06 &-2.53 &130\, \\ 
20 &-2.93 &-3.13 &38\, &70 &-2.04 &-2.11 &283\, \\ 
21 &-2.92 &-2.92 &57\, &\nullc{71} &\nullc{-2.04} &\nullc{-2.95} &\nullc{54}\, \\ 
22 &2.86 &3.26 &31\, &\nullc{72} &\nullc{-2.03} &\nullc{-3.08} &\nullc{40}\, \\ 
23 &-2.83 &-2.82 &74\, &73 &-2.02 &-2.30 &210\, \\ 
24 &2.82 &2.37 &180\, &\nullc{74} &\nullc{-2.01} &\nullc{-3.67} &\nullc{15}\, \\ 
25 &-2.81 &-2.13 &276\, &75 &2.00 &2.62 &109\, \\ 
\nullc{26} &\nullc{2.81} &\nullc{3.48} &\nullc{21}\, &76 &-1.98 &-2.38 &171\, \\ 
27 &-2.79 &-3.01 &47\, &77 &1.98 &1.43 &795\, \\ 
28 &2.70 &2.87 &64\, &78 &1.96 &1.69 &549\, \\ 
29 &-2.66 &-3.15 &37\, &79 &-1.95 &-1.47 &746\, \\ 
\nullc{30} &\nullc{-2.58} &\nullc{-3.85} &\nullc{11}\, &80 &1.95 &1.95 &361\, \\ 
31 &-2.56 &-2.84 &71\, &\nullc{81} &\nullc{1.95} &\nullc{2.81} &\nullc{77}\, \\ 
32 &-2.55 &-1.72 &524\, &82 &-1.94 &-1.41 &813\, \\ 
33 &-2.54 &-2.63 &106\, &\nullc{83} &\nullc{1.94} &\nullc{3.40} &\nullc{26}\, \\ 
34 &-2.54 &-2.69 &98\, &84 &1.94 &1.30 &948\, \\ 
35 &2.53 &2.30 &209\, &\nullc{85} &\nullc{-1.94} &\nullc{-3.27} &\nullc{30}\, \\ 
36 &2.48 &2.45 &148\, &86 &-1.93 &-1.11 &1190\, \\ 
37 &-2.47 &-2.29 &212\, &87 &-1.93 &-1.37 &872\, \\ 
38 &-2.46 &-3.21 &33\, &\nullc{88} &\nullc{-1.93} &\nullc{-3.44} &\nullc{24}\, \\ 
39 &-2.43 &-2.44 &154\, &\nullc{89} &\nullc{-1.92} &\nullc{-3.07} &\nullc{41}\, \\ 
40 &2.43 &2.71 &94\, &90 &-1.92 &-1.50 &726\, \\ 
41 &2.40 &2.86 &66\, &\nullc{91} &\nullc{1.90} &\nullc{3.62} &\nullc{16}\, \\ 
42 &-2.34 &-2.60 &115\, &\nullc{92} &\nullc{-1.90} &\nullc{-2.82} &\nullc{75}\, \\ 
43 &-2.34 &-2.98 &50\, &93 &1.89 &1.25 &1007\, \\ 
\nullc{44} &\nullc{-2.33} &\nullc{-3.80} &\nullc{12}\, &\nullc{94} &\nullc{1.87} &\nullc{3.89} &\nullc{8}\, \\ 
45 &-2.32 &-2.06 &306\, &\nullc{95} &\nullc{-1.86} &\nullc{-3.49} &\nullc{19}\, \\ 
46 &2.29 &1.81 &444\, &\nullc{96} &\nullc{-1.86} &\nullc{-2.08} &\nullc{300}\, \\ 
47 &-2.27 &-1.17 &1110\, &97 &1.85 &1.20 &1074\, \\ 
48 &2.26 &1.97 &347\, &\nullc{98} &\nullc{-1.83} &\nullc{-2.90} &\nullc{60}\, \\ 
\nullc{49} &\nullc{2.24} &\nullc{3.75} &\nullc{14}\, &99 &1.83 &1.39 &833\, \\ 
\nullc{50} &\nullc{2.20} &\nullc{3.88} &\nullc{9}\, &100 &-1.82 &-1.95 &367\, \\ 
\hline \ \ 
\end{tabular}
  \caption{Tellipsoid in action with Top 100 $\hat{u}^*_i$'s. Corresponding $t_i$'s and their rank are also shown. The results are for some $X$ realization from Fig.~\ref{fig.Ttl100R100nt10}. Tellipsoid = 22 NoFPs; raw $t$-statistics = 68 NoFPs. Truly null genes are printed in bold-sans typeface.}
  \label{tab.tiui}
\end{table}

Interestingly, with a smaller $n$, SAM outperforms  the other two techniques. This is not entirely surprising as a smaller $n$ can make the noise in the per gene pooled variance $s_i$ (and possibly the equivalent quantity in the EDGE algorithm) more prominent. Nevertheless, SAM does mitigate this issue in some measure by using the exchangeability factor $s_0$ to adjust the effective pooled variance~\citep{tusher2001sam}.

\subsection{Discussion}\label{subsec.discussion}

By allowing researchers  to examine the simultaneous expressions of enormous numbers of genes,  microarrays promised to revolutionize the understanding of complex diseases and usher in an era of personalized medicine. However, the shift in perception of that promise is palpable in the literature. A 1999 \textit{Nature Genetics} article \citep{lander1999ah} is entitled ``Array of hope," but a 2005 \textit{Nature Reviews} article \citep{frantz2005ap} is entitled ``An array of problems." It is not unusual for impacts of new technologies to be overestimated when first deployed, then to have the expectations moderated as the technologies reveal new complexities in the problems they are designed to solve.  In the study of microarray data, the need for exceeding care in the design and regularization of experiments and data collection are understood to be critical,  but the biggest hindrance to progress   has been the data interpretation. In particular, as \citet{efron2007cal} and \citet{owen2005vnf} point out, the biggest challenge seems to be the treatment of \emph{intrinsic} inter-gene correlation.

In most microarray data there are at least three vital resources: (i) identifiability (ii) immense parallel structure, and (iii) inter-gene correlation itself. Thoughtful analyses in the papers by Efron~\citep{efron2005bfa, efron2000rfc} have suggested this view of the rich information structure inherent in the data. In this light, Tellipsoid can be viewed as exploiting more than correlation as a means of sharing  information across tests, as it also involves identifiability.

A crucial step in  formulating Tellipsoid was the comprehension of the effects of  inter-gene correlation on $ \cov(t_i,t_\ip)$. In light of Observations 1--3, the choice of the Mahalanobis distance was intuitive, as it is already known to give computationally attractive solutions through the matrix inversion lemma.

Limited time and resources -- and perhaps also the necessity for scientific focus -- often require biomedical researchers to work on only a small number of ``hot (gene) prospects." Even under such highly conservative conditions, however,  misleading results can occur, as evident in the results of Figs.~\ref{fig.Ttl300}--\ref{fig.Ttl_100_nt10_sim}.  For all their careful development and statistical power,  even   state-of-the-art tools like EDGE and SAM can  report   spurious gene lists. The extra statistical power of Tellipsoid promises to further guard against anomalous results that can have serious consequences for the trajectory of a study of gene function, causation, and interaction.

\section{Conclusion}\label{sec.conclusion}
This paper  has reported the development and testing of a novel framework for the detection of differential gene expression. The framework combines the exploitation of inter-gene correlation to share information across tests, with identifiability -- the fact that in most microarray data sets, a large proportion of genes can be identified \textit{a priori} as non-differential. When applied to the widely used two-sample $t$-statistic approach, this viewpoint yielded an elegant differential analysis technique,  Tellipsoid, which requires as inputs only a gene expression matrix, related two-sample labels, and the size of desired gene-list $R$.  Tellipsoid was tested on the prostate cancer data of \citet{singh2002gec} and some simulated data. Compared to SAM~\citep{tusher2001sam}, EDGE~\citep{storey2007odp}, and the raw $t$-statistic approach itself, Tellipsoid shows substantial improvement in statistical power. Usually, with increase in microarray samples, power tends to increase considerably, but, even for small sample sizes, Tellipsoid's performance improvement is noticeable.  The software (coded in MATLAB and in R) and test data sets are available at \url{www.egr.msu.edu/~desaikey}.

\section*{Funding}
KD is supported by a graduate research fellowship from the \href{http://biomodel.msu.edu/}{Quantitative Biology Initiative} at Michigan State University.

\section*{Acknowledgement}
We are grateful to the anonymous reviewers of an earlier publication~\citep{desai2007dnf} for their insightful remarks that seeded this work. We also appreciate the support of the \href{http://www.hpcc.msu.edu}{Michigan State University High Performance Computing Center}.

\renewcommand{\theequation}{A-\arabic{equation}}
\setcounter{equation}{0}  

\section*{APPENDIX}  

Suppose that
\beq
\matr{\Sigma}^{-1} = \begin{pmatrix} \matr{A} & \matr{B} \\ \matr{C} & \matr{D} \end{pmatrix}\,\,\textrm{and}\,\, \tilde{\vect{u}}_{(1)} = \vect{t}_{(1)} - \vect{u}_{(1)}.
\label{eq.utilde}
\eeq
We also have $\matr{C} = \matr{B}\tr$.
Substituting these in Eqn.~\ref{eq.ustar} yields:
\beq
\tilde{\vect{u}}_{(1)}^* = \underset{\tilde{\vect{u}}_{(1)} \in \mathbb{R}^{m-c}}{\argmin} \vect{t}_{(0)}\tr \matr{A} \vect{t}_{(0)} + 2\tilde{\vect{u}}_{(1)}\tr \matr{C} \vect{t}_{(0)} + \tilde{\vect{u}}_{(1)}\tr\matr{D} \tilde{\vect{u}}_{(1)}.
\label{eq.simplified}
\eeq
In Eqn.~\ref{eq.simplified}, by setting the gradient w.r.t $\tilde{\vect{u}}_{(1)}$ to 0, we obtain:
\beq
\tilde{\vect{u}}_{(1)}^* = -\matr{C}\tr\matr{D}^{-1}\vect{t}_{(0)}.
\label{eq.tildeustar}
\eeq
Now for $\matr{\Sigma}^{-1}$, we can appeal to the matrix inversion lemma \citep{golub1996mc}:
\newcommand{\Sig}[1]{\mathbf{\Sigma}_{(#1)}}
\newcommand{\Qmat}{\mathbf{Q}}
\begin{align*}
\matr{\Sigma}^{-1}=\begin{pmatrix} \Sig{00}^{-1}\left(1+\Sig{01}\Qmat^{-1}\Sig{10}\Sig{00}^{-1}\right) & -\Sig{00}^{-1}\Sig{01}\Qmat^{-1} \\ -\Qmat^{-1}\Sig{10}\Sig{00}^{-1} & \Qmat^{-1} \end{pmatrix},
\end{align*}
where $\Qmat = \Sig{11}-\Sig{10}\Sig{00}^{-1}\Sig{01}$. Plugging this in Eqn.~\ref{eq.tildeustar} yields:
\beq
\tilde{\vect{u}}_{(1)}^* = \Sig{10}\Sig{00}^{-1}\vect{t}_{(0)}.
\label{eq.ustarttilde}
\eeq
Combining Eqn.~\ref{eq.ustarttilde} with Eqn.~\ref{eq.utilde} provides the desired expression:
\begin{equation*}
\vect{u}_{(1)}^* =  \vect{t}_{(1)} - \matr{\Sigma}_{(10)}\matr{\Sigma}_{(00)}^{-1}\vect{t}_{(0)} \qed.
\end{equation*}

\bibliographystyle{plainnat}
\bibliography{Tellipsoid_arXiv}

\begin{thebibliography}{32}
\providecommand{\natexlab}[1]{#1}
\providecommand{\url}[1]{\texttt{#1}}
\expandafter\ifx\csname urlstyle\endcsname\relax
  \providecommand{\doi}[1]{doi: #1}\else
  \providecommand{\doi}{doi: \begingroup \urlstyle{rm}\Url}\fi

\bibitem[Benjamini and Hochberg(1995)]{benjamini1995cfd}
Y.~Benjamini and Y.~Hochberg.
\newblock {Controlling the false discovery rate: a practical and powerful
  approach to multiple testing}.
\newblock \emph{Journal of the Royal Statistical Society. Series B.
  Methodological}, 57\penalty0 (1):\penalty0 289--300, 1995.

\bibitem[Cui et~al.(2005)Cui, Hwang, Qiu, Blades, and Churchill]{cui2005ist}
X.~Cui, J.T.G. Hwang, J.~Qiu, N.J. Blades, and G.A. Churchill.
\newblock {Improved statistical tests for differential gene expression by
  shrinking variance components estimates}.
\newblock \emph{Biostatistics}, 6\penalty0 (1):\penalty0 59--75, 2005.

\bibitem[Deller~Jr et~al.(1993)Deller~Jr, Proakis, and Hansen]{dellerjr1993dtp}
J.R. Deller~Jr, J.G. Proakis, and J.H. Hansen.
\newblock \emph{{Discrete Time Processing of Speech Signals}}.
\newblock Prentice Hall PTR Upper Saddle River, NJ, USA, 1993.

\bibitem[Desai et~al.(2007)Desai, Deller, and McCormick]{desai2007dnf}
K.~Desai, J.R. Deller, Jr., and J.J. McCormick.
\newblock The distribution of the number of false discoveries in highly
  correlated dna microarray data.
\newblock \emph{Submitted to the Annals of Applied Statstics, preprint at
  \url{http://www.egr.msu.edu/~desaikey}}, 2007.

\bibitem[Devijver and Kittler(1982)]{devijver1982prs}
P.A. Devijver and J.~Kittler.
\newblock \emph{{Pattern Recognition: A Statistical Approach}}.
\newblock Englewood Cliffs, New Jersey etc, 1982.

\bibitem[Dudoit et~al.(2002)Dudoit, Yang, Callow, and Speed]{dudoit2002smi}
S.~Dudoit, Y.H. Yang, M.J. Callow, and T.P. Speed.
\newblock {Statistical methods for identifying differentially expressed genes
  in replicated cDNA microarray experiments}.
\newblock \emph{Statistica Sinica}, 12\penalty0 (1):\penalty0 111--139, 2002.

\bibitem[Efron(2000)]{efron2000rfc}
B.~Efron.
\newblock {RA Fisher in the 21st Century}.
\newblock \emph{Statistics for the 21st Century: Methodologies for Applications
  of the Future}, 1:\penalty0 09, 2000.

\bibitem[Efron(2004)]{efron2004lss}
B.~Efron.
\newblock {Large-Scale Simultaneous Hypothesis Testing: The Choice of a Null
  Hypothesis.}
\newblock \emph{Journal of the American Statistical Association}, 99\penalty0
  (465):\penalty0 96--105, 2004.

\bibitem[Efron(2005)]{efron2005bfa}
B.~Efron.
\newblock {Bayesians, Frequentists, and Scientists.}
\newblock \emph{Journal of the American Statistical Association}, 100\penalty0
  (469):\penalty0 1--6, 2005.

\bibitem[Efron(2006)]{efron2006mem}
B.~Efron.
\newblock {Microarrays, Empirical Bayes, and the Two-Groups Model}.
\newblock \emph{Preprint, Dept. of Statistics, Stanford University}, 2006.

\bibitem[Efron(2007)]{efron2007cal}
B.~Efron.
\newblock {Correlation and Large-Scale Simultaneous Significance Testing}.
\newblock \emph{Journal of the American Statistical Association}, 102\penalty0
  (477):\penalty0 93--103, 2007.

\bibitem[Efron et~al.(2001)Efron, Tibshirani, Storey, and Tusher]{efron2001eba}
B.~Efron, R.~Tibshirani, J.D. Storey, and V.~Tusher.
\newblock {Empirical Bayes Analysis of a Microarray Experiment.}
\newblock \emph{Journal of the American Statistical Association}, 96\penalty0
  (456):\penalty0 1151--1161, 2001.

\bibitem[Frantz(2005)]{frantz2005ap}
S.~Frantz.
\newblock {An array of problems.}
\newblock \emph{Nat Rev Drug Discov}, 4\penalty0 (5):\penalty0 362--3, 2005.

\bibitem[Golub and Van~Loan(1996)]{golub1996mc}
G.H. Golub and C.F. Van~Loan.
\newblock \emph{{Matrix Computations}}.
\newblock Johns Hopkins University Press, 1996.

\bibitem[Hedenfalk et~al.(2001)Hedenfalk, Duggan, Chen,
  et~al.]{hedenfalk2001gep}
I.~Hedenfalk, D.~Duggan, Y.~Chen, et~al.
\newblock {Gene-Expression Profiles in Hereditary Breast Cancer}.
\newblock \emph{New England Journal of Medicine}, 344\penalty0 (8):\penalty0
  539--548, 2001.

\bibitem[Irizarry et~al.(2003)Irizarry, Bolstad, Collin, Cope, Hobbs, and
  Speed]{irizarry2003sag}
R.A. Irizarry, B.M. Bolstad, F.~Collin, L.M. Cope, B.~Hobbs, and T.P. Speed.
\newblock {Summaries of Affymetrix GeneChip probe level data}.
\newblock \emph{Nucleic Acids Research}, 31\penalty0 (4):\penalty0 e15, 2003.

\bibitem[Kerr et~al.(2000)Kerr, Martin, and Churchill]{kerr2000avg}
M.K. Kerr, M.~Martin, and G.A. Churchill.
\newblock {Analysis of Variance for Gene Expression Microarray Data}.
\newblock \emph{Journal of Computational Biology}, 7\penalty0 (6):\penalty0
  819--837, 2000.

\bibitem[Lander(1999)]{lander1999ah}
E.S. Lander.
\newblock {Array of hope}.
\newblock \emph{Nature Genetics}, 21\penalty0 (1):\penalty0 3--4, 1999.

\bibitem[Langaas et~al.(2005)Langaas, Ferkingstad, and
  Lindqvist]{langaas2005ept}
M.~Langaas, E.~Ferkingstad, and B.H. Lindqvist.
\newblock {Estimating the proportion of true null hypotheses, with application
  to DNA microarray data}.
\newblock \emph{J Roy Stat Soc, Series B}, 67:\penalty0 555--72, 2005.

\bibitem[Lee et~al.(2000)Lee, Kuo, Whitmore, and Sklar]{lee2000irm}
M.L.T. Lee, F.C. Kuo, GA~Whitmore, and J.~Sklar.
\newblock {Importance of replication in microarray gene expression studies:
  Statistical methods and evidence from repetitive cDNA hybridizations}.
\newblock \emph{Proceedings of the National Academy of Sciences of the United
  States of America}, 97\penalty0 (18):\penalty0 9834, 2000.

\bibitem[Leek et~al.(2006)Leek, Monsen, Dabney, and Storey]{leek2006eea}
J.T. Leek, E.~Monsen, A.R. Dabney, and J.D. Storey.
\newblock {EDGE: extraction and analysis of differential gene expression},
  2006.

\bibitem[Lonnstedt and Speed(2002)]{lonnstedt2002rmd}
I.~Lonnstedt and T.~Speed.
\newblock {Replicated microarray data}.
\newblock \emph{Statistica Sinica}, 12\penalty0 (1):\penalty0 31--46, 2002.

\bibitem[Mahalanobis(1936)]{mahalanobis1936gds}
P.C. Mahalanobis.
\newblock {On the generalized distance in statistics}.
\newblock \emph{Proc Natl Inst Sci India}, 2\penalty0 (1):\penalty0 49--55,
  1936.

\bibitem[Newton et~al.(2001)Newton, Kendziorski, Richmond, Blattner, and
  Tsui]{newton2001dve}
MA~Newton, CM~Kendziorski, CS~Richmond, F.R. Blattner, and KW~Tsui.
\newblock {On Differential Variability of Expression Ratios: Improving
  Statistical Inference about Gene Expression Changes from Microarray Data}.
\newblock \emph{Journal of Computational Biology}, 8\penalty0 (1):\penalty0
  37--52, 2001.

\bibitem[Owen(2005)]{owen2005vnf}
A.B. Owen.
\newblock {Variance of the number of false discoveries}.
\newblock \emph{Journal of the Royal Statistical Society Series B}, 67\penalty0
  (3):\penalty0 411--426, 2005.

\bibitem[Pawitan et~al.(2005)Pawitan, Murthy, Michiels, Ploner, and
  Journals]{pawitan2005bef}
Y.~Pawitan, K.R.K. Murthy, S.~Michiels, A.~Ploner, and O.~Journals.
\newblock {Bias in the estimation of false discovery rate in microarray
  studies}.
\newblock \emph{Bioinformatics}, 21\penalty0 (20):\penalty0 3865--3872, 2005.

\bibitem[Singh et~al.(2002)Singh, Febbo, Ross, et~al.]{singh2002gec}
D.~Singh, P.G. Febbo, K.~Ross, et~al.
\newblock {Gene expression correlates of clinical prostate cancer behavior}.
\newblock \emph{Cancer Cell}, 1\penalty0 (2):\penalty0 203--209, 2002.

\bibitem[Storey(2002)]{storey2002daf}
J.D. Storey.
\newblock {A direct approach to false discovery rates}.
\newblock \emph{Journal of the Royal Statistical Society Series B(Statistical
  Methodology)}, 64\penalty0 (3):\penalty0 479--498, 2002.

\bibitem[Storey et~al.(2007)Storey, Dai, and Leek]{storey2007odp}
J.D. Storey, J.Y. Dai, and J.T. Leek.
\newblock {The optimal discovery procedure for large-scale significance
  testing, with applications to comparative microarray experiments}.
\newblock \emph{Biostatistics}, 8\penalty0 (2):\penalty0 414, 2007.

\bibitem[Tibshirani and Wasserman(2006)]{tibshirani2006csd}
R.~Tibshirani and L.~Wasserman.
\newblock {Correlation-sharing for detection of differential gene expression}.
\newblock \emph{Arxiv preprint math.ST/0608061}, 2006.

\bibitem[Tsai et~al.(2003)Tsai, Chen, Chen, and Journals]{tsai2003tde}
C.A. Tsai, Y.J. Chen, J.J. Chen, and O.~Journals.
\newblock {Testing for differentially expressed genes with microarray data}.
\newblock \emph{Nucleic Acids Research}, 31\penalty0 (9):\penalty0 e52, 2003.

\bibitem[Tusher et~al.(2001)Tusher, Tibshirani, and Chu]{tusher2001sam}
V.~Tusher, R.~Tibshirani, and C~Chu.
\newblock Significance analysis of microarrays applied to ionizing radiation
  response.
\newblock \emph{Proceedings of the National Academy of Sciences}, 98:\penalty0
  5116--5121, 2001.

\end{thebibliography}
\end{document}